\documentclass[12pt,preprint]{aastex}
\usepackage{apjfonts}
\usepackage{latexsym}
\shorttitle{Variable stars in Leo I}
\shortauthors{Fiorentino et al.}
\begin{document}
\def\gsim{\;\lower.6ex\hbox{$\sim$}\kern-6.7pt\raise.4ex\hbox{$>$}\;}
\def\lsim{\;\lower.6ex\hbox{$\sim$}\kern-6.7pt\raise.4ex\hbox{$<$}\;}
\title{On the central helium--burning variable stars of the LeoI dwarf spheroidal galaxy}

\author{G. Fiorentino$^{1}$, P. B. Stetson$^{2}$, M. Monelli$^{3,4}$,
  G. Bono$^{5,6}$, E. J. Bernard$^{7}$ and A. Pietrinferni$^{8}$}
\email{giuliana.fiorentino@oabo.inaf.it}
\affil{$^{1}$INAF-Osservatorio Astronomico di Bologna, via Ranzani 1, 40127, Bologna.}
\affil{$^{2}$National Research Council, 5071 West Saanich Road, Victoria, BC V9E 2E7, Canada.}
\affil{$^{3}$Instituto de Astrof\'{i}sica de Canarias, Calle Via Lactea s/n, E38205 La Laguna, Tenerife, Spain.}
\affil{$^{4}$Departmento de Astrof\'{i}sica, Universidad de La Laguna, E38200 La Laguna, Tenerife, Spain.}
\affil{$^{5}$Dipartimento di Fisica, Universit\'{a} di Roma Tor Vergata, Via della Ricerca Scientifica 1, 00133 Roma, Italy.}
\affil{$^{6}$INAF-Osservatorio Astronomico di Roma, Via Frascati 33, 00040 Monte Porzio Catone, Italy}
\affil{$^{7}$SUPA, Institute for Astronomy, University of Edinburgh, Royal Observatory, Blackford Hill, Edinburgh EH9 3HJ, UK.}
\affil{$^{8}$INAF-Osservatorio Astronomico di Teramo, via M. Maggini, 64100, Teramo.}

\begin{abstract}
We present a study of short period, central helium--burning variable
stars in the Local Group dwarf spheroidal galaxy LeoI, including 106 RR Lyrae
stars and 51 Cepheids. So far, this is the largest sample of Cepheids
and the largest Cepheids to RR Lyrae ratio found in such a kind of galaxy. The comparison with other Local Group dwarf spheroidals, Carina and Fornax, shows that the period distribution of RR
Lyrae stars is quite similar, suggesting similar properties of the
parent populations, whereas the Cepheid period distribution in LeoI
peaks at longer periods (P$\sim$ 1.26d instead of $\sim$0.5d) and spans over a
broader range, from 0.5 to 1.78d.
Evolutionary and pulsation predictions indicate, assuming a mean
metallicity peaked within --1.5$\lsim$[Fe/H]$\lsim$--1.3, that the 
current sample of LeoI Cepheids traces a unique mix of Anomalous Cepheids 
(blue extent of the red--clump, partially electron degenerate central helium-burning
stars) and short-period classical Cepheids (blue--loop, quiescent 
central helium-burning stars). Current evolutionary prescriptions also
indicate that the transition mass between the two different groups of
stars is M$_{HeF}\sim$2.1~M$_{\odot}$, and it is constant
for stars metal--poorer than [Fe/H]$\sim$--0.7.  
Finally, we briefly outline the different implications of the 
current findings on the star formation history of LeoI.
\end{abstract}


\section{Introduction} 
Resolved stellar populations in nearby galaxies are a fundamental laboratory 
to constrain the impact of the environment on stellar evolution. The observables 
adopted to investigate these stellar systems typically rely on
color-magnitude diagrams (CMDs), on kinematics or on chemical abundances. 
Pulsation properties of variable stars can also be adopted to provide firm 
constraints on the age, on the metallicity distribution of the parent 
stellar populations (e.g., Fornax, 
\citealt{feast12}), and on the star formation episodes experienced 
by these interesting stellar systems. This unique opportunity becomes even 
more important for stellar
systems for which current ground-based and space telescopes do not
allow us to firmly identify the main sequence turn-off stars of the
different star formation episodes (e.g., CentaurusA,
\citealt{rejkuba05}; M32, \citealt{fiorentino12a}). 

The Local Group dwarf spheroidal (dSph) galaxy LeoI plays a key role 
in this context. LeoI was discovered more than half century ago 
\citep{harrington50} and \citet{lee93} suggested that its stellar content 
is younger than typical nearby dSphs: in particular, it is dominated by 
an intermediate-age population of $\sim$3 Gyr. This hypothesis was 
soundly supported by \citet{gallart99}, who investigated the star 
formation history (SFH) using deep optical CMDs reaching the oldest 
main sequence turn-off. Assuming [Fe/H]=--1.7, they found that
LeoI experienced significant star formation events between
approximately 7 and 1 Gyr ago. The presence in LeoI of an old
stellar population ($\sim$10 Gyr), was demonstrated  
by \citet{held00} who first unambiguously identified a well populated 
horizontal branch (HB, low-mass central helium-burning stars),
further confirmed by the subsequent identification of 74 candidate 
RR Lyrae stars \citep[hereinafter RRLs][]{held01}. The same authors 
provided pulsation properties for 54 of them. 
The mean metallicity 
of LeoI has been estimated using the infrared 
Ca-triplet of red giant branch (RGB) stars. 
By using two independent calibrations rooted on Galactic 
globular clusters and on the metallicity scale provided by 
\citet{carretta09}, \citet{bosler07} found a mean metallicity of 
[Fe/H]$\sim$--1.34 ($\sigma$=0.21 dex) using 101 stars and 
\citet{gullieuszik09} found [Fe/H]$\sim$--1.37 ($\sigma$=0.18 dex) using 50
stars. This is in quite good agreement with recent medium-resolution spectroscopy
 of 850 RGB stars that provided a mean metallicity of
 [Fe/H]$\sim$--1.43 \citep[$\sigma$=0.33 dex,][]{kirby11}. It is worth mentioning that LeoI seems to have a broad metallicity 
distribution, iron abundances range from $\sim$--2.15 to 
$\sim$--1. A large spread in metallicity appears to be 
typical of several nearby dSphs. However, recent spectroscopic 
measurements based on high-resolution spectra indicate a significant 
decrease when compared with estimates based on the Ca-triplet index \citep{fabrizio12}.

Interestingly enough, \citet{hodge78} suggested the existence of a
large sample of Anomalous Cepheids (ACs) in LeoI. They identified 
17 AC candidates, but the periods were estimated only for 5 of them. 
This finding seems partially at odds with the metal--rich chemical 
composition suggested by the above spectroscopic analysis, since theory and 
observations indicate that ACs are associated with more metal--poor 
stellar populations ([Fe/H] $\lsim$--1.5).  
In this letter, we discuss the discovery of new RRLs and Cepheids. 
We focus our attention on the proper classification of the latter group by 
using pulsation and evolutionary models, and on its consequences on the SFH of
LeoI.

\begin{figure} \includegraphics[width=8.5cm]{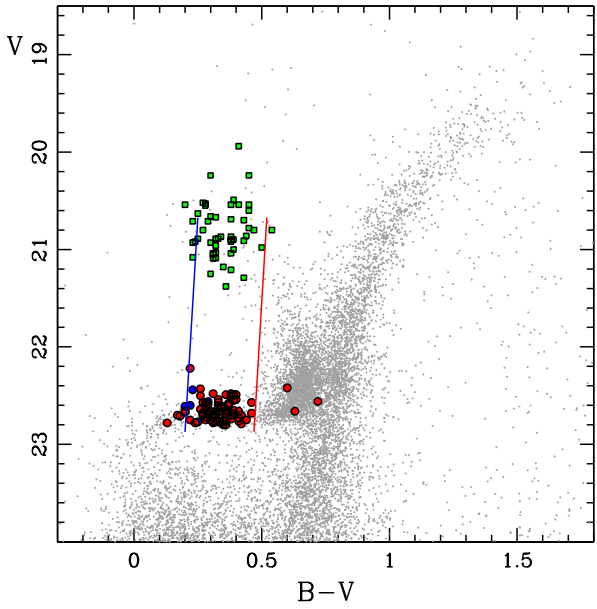} 
\caption{The $V$, $B-V$ Color-Magnitude diagram of LeoI. Red and blue circles 
display fundamental and first overtone RRLs. Very red variables (B-V > 0.5 mag) are very likely blends having the amplitude in $B$ smaller than in $V$ band. 
Squares mark Cepheids. The solid vertical lines display the predicted instability 
strip for RRLs and Cepheids \citep{fiorentino06}.
\label{fig:cmd}} 
\end{figure}

\section{Data samples and data reduction}\label{data} 
This investigation is based on a large number of optical images 
collected from different ground-based facilities. The data reduction 
and calibration is part of the effort of one of us (P.B.S.) to maintain 
a data base of homogeneous photometry of resolved stellar clusters
and galaxies. The LeoI catalog includes $B$, $V$, $R$, $I$ images from 
29 observing runs on 16 different telescopes during the period from 1983 January 
to 2002 November. A full description of the data, including the methods used to 
identify candidate variable stars \citep{welch93} and to
determine the pulsational properties \citep{stetson98,fiorentino10a}, 
as well as their light curves, will be presented in a forthcoming
paper. Fig.\ref{fig:cmd} presents a ($V$, $B - V$) color-magnitude diagram (CMD) 
containing 8588 stars located within a box of $\sim$40 square arcmin centered on LeoI. 
The entire sample spans from the tip of the RGB ($V \sim$ 19.5 mag)
down to $\sim 0.5$ mag below the old main sequence turn-off. Here, we
show a zoom on the HB, populated by old stars 
($>$10 Gyr) and the red clump (RC), typical of intermediate-age stars.
Note that the RC extends to brighter magnitudes and bluer colors,
typical of not-so-metal--rich environments ([Fe/H] $\le$--1.3).
The strong intermediate-age population \citep{gallart99} is also evident in the blue
plume of stars that reaches the magnitude level of the HB.

\section{Analysis of variable stars}\label{variable}

We have identified 157 strong candidate variable stars, and they are plotted in 
Fig.\ref{fig:cmd}. The number of phase points per object ranges from 20 ($B$) 
to more than 100 ($V$). We estimated accurate periods and mean magnitudes for 
138 variable stars.    

Two groups of variable stars with similar colors, 0.2 $<B$--$V<$ 0.5
mag, can be easily identified: a fainter one distributed along the HB 
($V \sim$ 22.5 mag) characterized by old, low-mass stars burning helium 
in an electron-degenerate core (circles) and a second one that is 
1.5--2.5 mag brighter (squares).  The instability strip (IS) predicted by 
pulsation models \citep[see vertical lines in Fig.\ref{fig:cmd},][]{fiorentino06} 
agrees quite well with both groups of variable stars. The former group (87 stars), 
according to their magnitude and color distributions can be safely identified 
as RRLs. The physical characterization of the latter group 
(51 stars) is less trivial, since this is a degenerate region of the
CMD.

In this magnitude range, intermediate-mass ($\gsim$ 1.5 M$_{\odot}$) 
stars burn helium in the core either in partially electron-degenerate 
conditions or quiescently. The latter group evolves along the 
so-called blue loop phase and in crossing the IS
gives the classical Cepheids (CCs), while the former one evolves off  
the bright extent of RC stars and in 
crossing the IS gives the ACs. In passing, we note 
that also old, low-mass population II Cepheids (P2C) can occupy the same
region in the CMD.  However, their pulsation properties allow us to
disentangle them from intermediate-mass stars \citep[P2C are 
characterized, at fixed luminosity, by longer periods, see also][]{monelli12b,fiorentino12c}.

\begin{figure*}
\includegraphics[width=9cm]{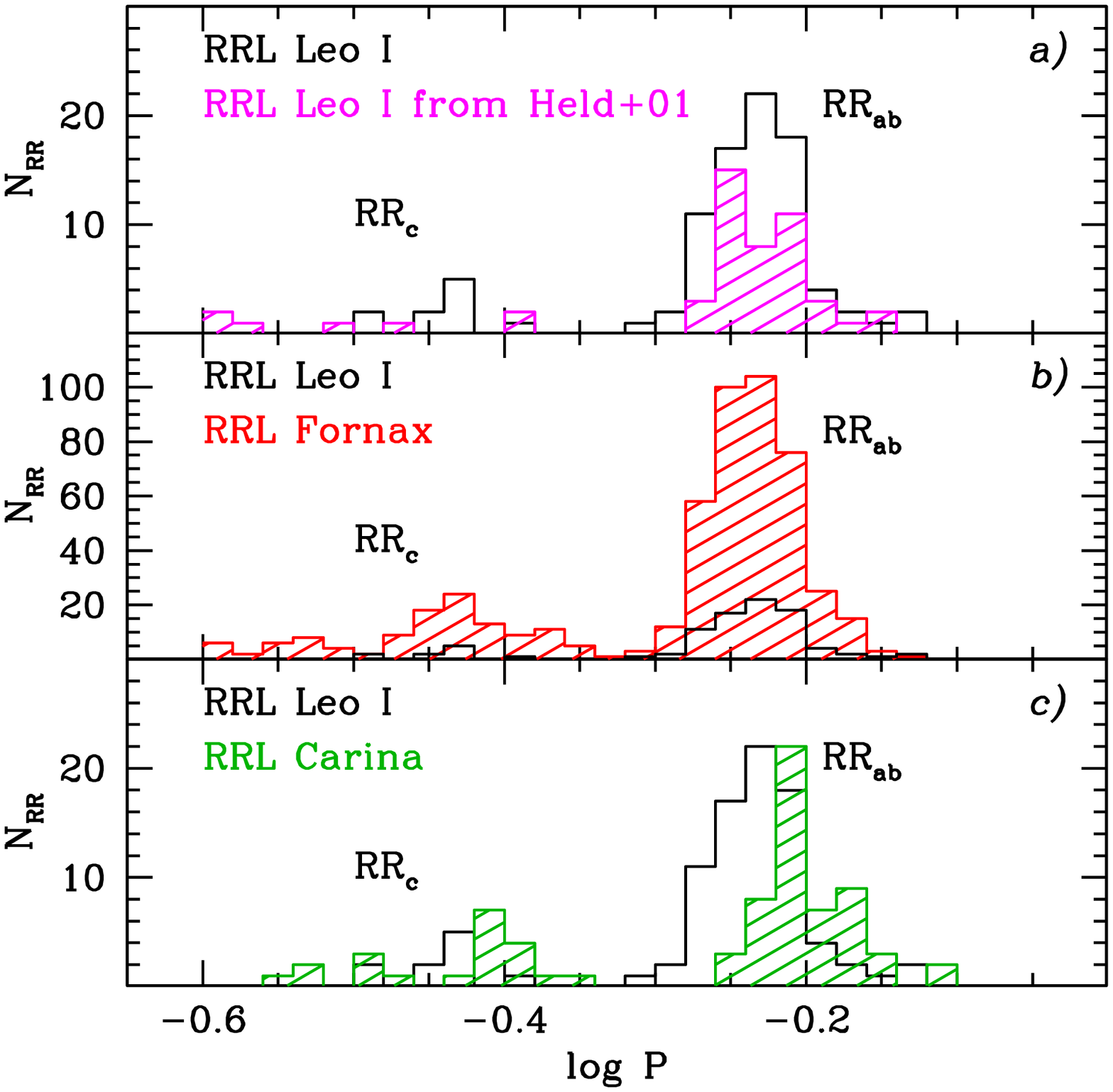}
\includegraphics[width=9cm]{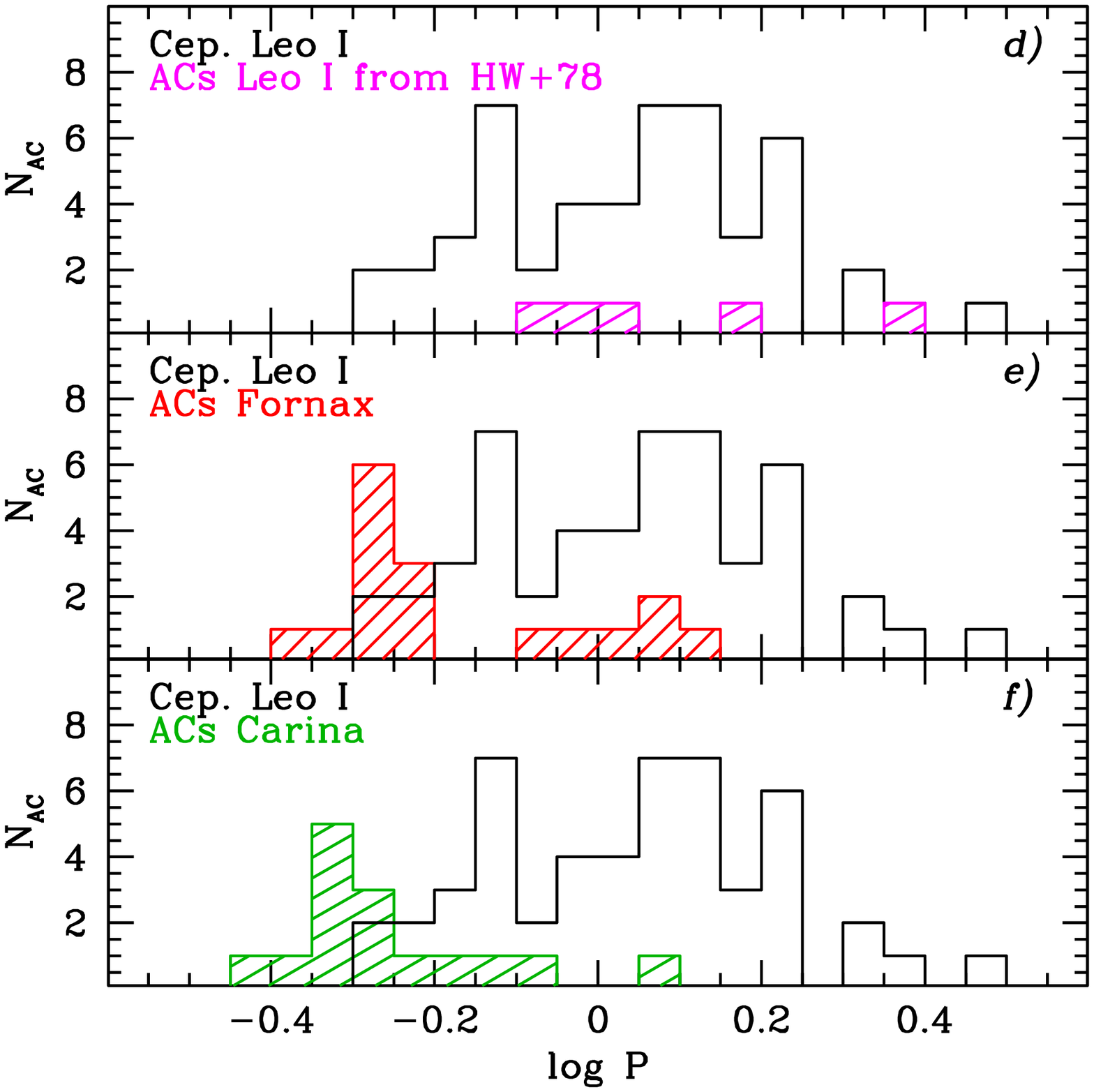}
\caption{Left -- period distribution of the RRL sample (90) 
identified in LeoI. The top panel shows the 
comparison with the period distribution of LeoI RRLs 
identified by \citet[54 out of 74 with well defined periods,][]{held01}, 
while the middle and the bottom panels display the comparison with 
RRLs in Fornax \citep[515,][]{bersier02} 
and in Carina \citep[77,][]{dallora03}, respectively. 
Right -- same as the left panels, but for ACs. From top to bottom the panels 
show the comparison between the period distribution of the current sample 
of 51 LeoI Cepheids with the 5 bona-fide ACs identified by \citet{hodge78} and with 
the ACs identified in Fornax \citep[17,][]{bersier02} and in Carina 
\citep[15,][]{dallora03}.   
\label{fig:per}}
\end{figure*}

\subsection{Period distribution}\label{per}

The period distributions of RRLs and Cepheids in LeoI are shown
  in Fig.\ref{fig:per}, panels {\it a)} and {\it d)} respectively.
The period distribution of LeoI RRLs shows a clear separation between the 77 fundamental mode
(RR$_{ab}$) and the nine first-overtone (RR$_c$) variables. The mean period of RRLs is 
an important parameter, since globular clusters hosting a 
sizable sample of RRLs show the so-called Oosterhoff dichotomy (Oosterhoff type I [OoI],
<P$_{ab}$>$\sim$0.55~d; Oosterhoff type II [OoII], <P$_{ab}$>$\sim$0.64~d). The mean period 
showed by RR$_{ab}$ variables in LeoI is <P$_{ab}$>= 0.59$\pm$0.05~d. 
The inclusion of RR$_c$ variables minimally affects the mean period, and indeed 
once we fundamentalize their periods we find <P>= 0.58$\pm$0.06~d. 
This suggests that LeoI is an Oosterhoff intermediate system,
supporting previous results \citep{held01}. The Oosterhoff dichotomy also shows up in the ratio between 
RR$_c$ and the total number of RRLs: OoI clusters show 
a ratio $\sim$0.2, while for OoII clusters this 
is equal to $\sim$0.5. Aware that low-amplitude stars may suffer for lower completeness
than high-amplitude stars, for LeoI we found $N_c$/($N_c$+$N_{ab}$)$\sim$0.10, 
thus suggesting that it might be more similar to an 
OoI cluster. These two features are in common with other nearby stellar systems 
\citep[see Fig.~2 in][]{bono03b} and further support the impact of the environment 
on pulsation properties of RRLs \citep{monelli12b}. The period distribution of LeoI Cepheids covers a broad range 
($-0.3$$\le \log P \le$$0.5$) with a wide peak at $\log P\sim$0.1. 
This is a period range that is poorly sampled in short-period variables 
due to the one-day alias. However, the very long time interval and the 
high number of phase points adopted in the 
current analysis allowed us to overcome this thorny problem. The histograms 
plotted in panel {\it d)} disclose the relevant improvement in the 
current sample of ACs when compared with the 17 previously identified ACs
\citep[only 5 with accurate periods;][]{hodge78}.
To further constrain the pulsation properties of RRLs and Cepheids in LeoI,  
we adopted two empirical calibrators, namely Carina and Fornax. We selected 
these dSphs because they are nearby and they host sizable samples of both 
RRLs and ACs.  

The variable star population of the Carina dSph is well understood. The
52 fundamental mode RRLs are peaked around P$=$0.631~d (see
Fig.\ref{fig:per}, panel {\it c)}), resembling
an OoII system \citep{dallora03} and are representative of the
oldest population ($\sim$ 12~Gyr). On the other hand, Carina appears to be an OoI
system according to $N_c$/($N_c$+$N_{ab}$)$\sim$0.20.
This evidence suggests that the pulsation properties of RRLs in LeoI 
and in Carina are relatively similar. Carina also hosts a genuine population of 15 ACs
\citep[see Fig.\ref{fig:per}, panel~ {\it f)};][]{dallora03},
which is fully in agreement with the observed intermediate-age stellar
population \citep[$\sim$ 600~Myr;][]{monelli03} and with a spectroscopic 
mean metallicity peaked around [Fe/H]=-1.80
\citep[$\sigma $= 0.24 dex,][]{fabrizio12}. The period distribution of 
Carina ACs peaks at periods that are systematically shorter than LeoI 
Cepheids. This indicates that the latter group might be the progeny  
of a different stellar population.\par
We still lack a complete census of evolved variable stars in Fornax. 
However, \citet{bersier02} reported a large number of RRLs (more than
500) suggesting a major star formation episode occurred at early epochs. 
The peak of RR$_{ab}$ periods is typical of an Oo intermediate, <P$_{ab}$>=
0.585. According to the fraction of $RR_c$ stars, $N_c$/($N_c$+$N_{ab}$)$\sim$0.23, Fornax
resembles an OoI group. These findings and the histograms plotted in the
panel {\it b)} of Fig.\ref{fig:per} disclose that RRLs in Fornax
and in LeoI also have similar pulsation properties. \citet{bersier02} also detected 
17 ACs showing a bi-modal period distribution, see panel {\it e)} of Fig.\ref{fig:per}. 
The short period peak is similar to the Carina peak, while the long period one 
is similar to the LeoI peak. No firm conclusion can be reached concerning the
difference between ACs in Fornax and Cepheids in LeoI. Indeed, the small 
number of Fornax ACs in the long period peak might be the consequence of both a
poor sampling and of the one-day aliasing.

The above empirical evidence indicates that the RRLs in Carina, Fornax and LeoI 
display similar pulsation properties, thus suggesting that they are the 
progeny of homogeneous old stellar populations. On the other hand, the current 
findings seem to suggest that a fraction of LeoI Cepheids could have 
different progenitors when compared with ACs in Carina and probably in Fornax.
This could be related to the different SFHs of the dSphs under study,
or simply, to observational biases that may hide a more cospicuous
presence of Cepheids in such stellar systems. Only more complete
studies, as the one presented here, will shed light on the reasons of such huge
difference in the Cepheid sample size of LeoI when compared with other
dSphs.

\section{Discussion and final remarks}\label{wes}

\begin{figure}
\includegraphics[width=9cm]{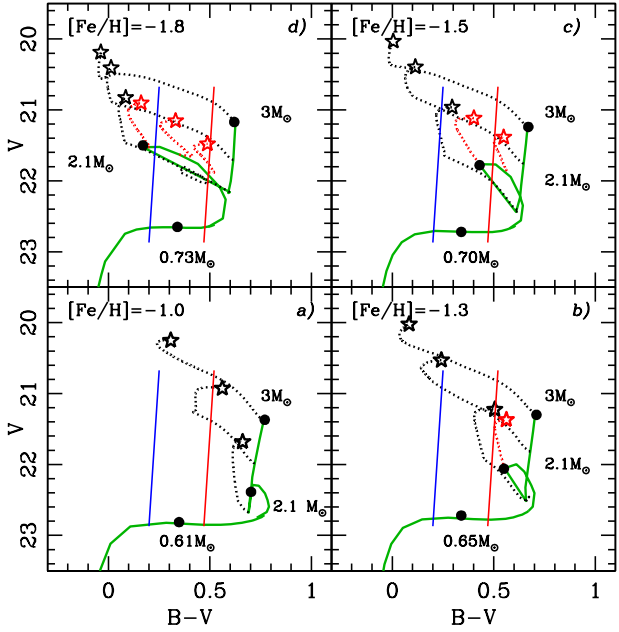}
\caption{Theoretical prediction for central helium-burning structures, based on scaled--solar 
evolutionary models from the BaSTI database \citep{pietrinferni04}, for the labeled values of
metallicity with a fixed $\Delta Y/ \Delta Z =$1.4 and a primordial
Helium Y=0.23. We assumed $\mu_0$=22.11 mag and E(B-V)=0.02 mag. Blue and red solid
  lines represents the theoretical IS, as in Fig.\ref{fig:cmd}. Green solid lines 
display the Zero Age helium-burning loci for stars with masses from $\sim$0.5 to 3 M$_{\odot}$.
Red and black dotted lines show evolutionary tracks during partially electron
degenerate or quiescent central helium-burning, respectively. The corresponding
red and black stars mark the approach to central 
helium exhaustion (10\% of the initial abundance). Black dotted lines
are for a 2.2, 2.6 and 3.0 M$_{\odot}$. Red dotted lines are different for [Fe/H]=--1.8 (2.1, 1.9, 1.7 M$_{\odot}$), 
[Fe/H]=--1.5 (2.1, 1.9 M$_{\odot}$),[Fe/H]=--1.3 (2.1 M$_{\odot}$). At each fixed chemical composition, 
the three black filled circles mark the stellar mass typical of RRLs (faintest point), the 
transition mass (M$_{HeF}$) between electron degenerate and quiescent central helium-burning
(middle) and the highest adopted stellar mass (brightest).
\label{fig:evo}}
\end{figure}

The 51 Cepheids detected in LeoI cover more than one visual magnitude in luminosity
(see Fig.\ref{fig:cmd}) and a broad range in periods ($\sim$ --0.3 $<$ Log P $<$
0.5, see Fig.\ref{fig:per}). This luminosity and color range ($M_V$$\lsim$--0.8
mag and 0.2 $\lsim B-V \lsim$ 0.5) is poorly 
investigated, because it can only be observed in metal--poor environments 
that experienced recent episodes of star formation. Precisely, only
stars with [Fe/H] $\lsim$--1.3 and masses around the
transition mass value M$_{HeF}$ ($\sim$2.1 M$_{\odot}$) populate this region of
the CMD. We recall that M$_{HeF}$ is defined as the transition mass between
star that ignite helium in the center under partial degenerate condition
(smaller masses) or in a quiescent way (larger masses). The transition mass
value is independent on the metallicity for [Fe/H] $\le$ --0.7. Since the pulsation
properties of the corresponding pulsators (ACs or CCs) are very
similar \citep{caputo04}, to properly identify the nature of our sample, 
a detailed comparison between evolutionary models and observations is
required. The panels in Fig.\ref{fig:evo} show scaled--solar evolutionary prescriptions for 
central helium-burning phase for masses from $\sim$ 0.5 to 3
M$_{\odot}$ covering a broad range of chemical compositions (see
labeled values). 

A glance at the evolutionary and pulsation predictions plotted in these panels 
reveals the strong sensitivity of intermediate-mass stars
to the chemical composition.
In particular, we note that the central helium burning loci of
structures with mass equal to the transition mass
M$_{HeF}$ (black dots in Fig.\ref{fig:evo}) 
become redder and fainter for increasing metallicity. This mass value falls
in the IS for a metallicity [Fe/H]$=$--1.5. 
In the following a detailed description of each panel:\\
 
\noindent {\bf --Panel} {\it a)}: [Fe/H]$\sim$--1 is the maximum metallicity that allows stars entering the
IS in the observed luminosity range. Only CCs with M$\ge$ 2.6 M$_{\odot}$ are expected
because smaller masses during their central helium-burning attain colors that are systematically redder than 
the IS, see Fig.\ref{fig:evo}.\par
\noindent {\bf --Panel} {\it b)}: A decrease in metallicity of 0.3
dex slightly affects the evolutionary scenario, stars burning helium in 
partially electron degenerate core do not (or slightly) cross the IS during their evolution.
The minimum mass value of quiescent central helium-burning structures crossing the IS decreases to 
2.2M$_{\odot}$, larger than M$_{HeF}$. This means that the system still produces mostly CCs, but
with periods shorter than in the previous case.\par 
\noindent {\bf --Panel} {\it c)}: A further decrease in the metallicity of 0.2 dex causes a dramatic change in 
the evolutionary scenario, and indeed the M$_{HeF}$ evolutionary model
falls inside the IS, within a magnitude range 
of V$\sim$21--22 mag. Thus, this region is almost entirely populated by ACs, while the brighter portion contains some CCs.\par    
\noindent {\bf --Panel} {\it d)}: The change in the evolutionary scenario outlined for [Fe/H]=--1.5 becomes even more   
pronounced for [Fe/H]=--1.8. The further decrease in metallicity implies that a 
significant portion of the hook performed by the central helium burning
loci for stellar masses less than M$_{HeF}$ falls inside the IS. This means that more metal--poor systems 
are expected to produce a significant number of ACs. Note that in this
metal--poor scenario ([Fe/H]$\lsim$--1.5) the 
occurrence of quiescent central helium-burning structures would be traced not 
only by short-period CCs, but also by Blue Tip stars, located at colors systematically bluer than Cepheids and marking the 
maximum extent in color of the blue loop where stars spend most of
their central helium-burning time ($20\lsim$V$\lsim 21$, B-V$\sim$0).\\    
%

\begin{figure}
\includegraphics[width=9cm]{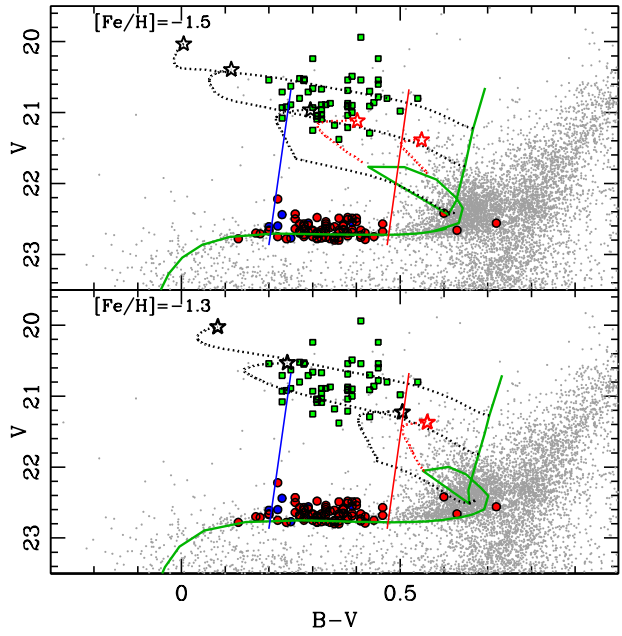}
\caption{Same as in Fig.\ref{fig:evo}, but for two metal--intermediate chemical compositions
namely [Fe/H]=--1.5 (top) and [Fe/H]=--1.3 (bottom). Symbols 
and colors used of the observed variables as in Fig.\ref{fig:cmd}.  
\label{fig:comp}}
\end{figure}

When we compare our observations with these theoretical prescriptions (see
Fig.\ref{fig:comp}), we
find that they agree quite well with the two intermediate chemical compositions, namely [Fe/H]$\sim$--1.5 (top) and 
[Fe/H]$\sim$--1.3 (bottom). Indeed, current theoretical predictions
for the central helium burning loci account for both low- (HB) and intermediate-mass (RC) 
helium-burning stars. In the following 
we outline the properties of LeoI Cepheids at fixed metallicity in
the current evolutionary scenario.  
If we assume that the mean metallicity of LeoI peaks at [Fe/H]$\sim$--1.3 
then the bulk of the LeoI Cepheids is made by short period CCs, with
masses systematically larger than M$_{HeF}$ and an age of their progenitors younger 
than 650 Myr. On the other hand, if we assume [Fe/H]$\sim$--1.5 then part of the LeoI Cepheids is made 
by ACs, with a typical star mass populating the IS equal to M$_{HeF}$. 
The interesting feature disclosed by the top 
panel of Fig.\ref{fig:comp} is that the bright tail of LeoI Cepheids seems  
to consist of CCs (quiescent central helium-burning) while the 
fainter tail seems to consist of ACs (partial electron degeneracy 
central helium-burning). This means that the stellar mass distribution 
of LeoI Cepheids should be larger and range from 2M$_{\odot}$ (700 Myr) 
to $\sim$3M$_{\odot}$ (250 Myr). 

Note that the above discussion is based on the assumption that the 
bulk of the bright variable stars is characterized by one mean metallicity. 
However, spectroscopic ([Fe/H]$\sim$ -1.43 $\pm$ 0.33; \citealt{kirby11})
and photometric indicators (width in color of the RGB) support the evidence that  
stellar populations in LeoI are characterized by a spread in
metallicity, thus suggesting that they might be a unique mix of both ACs
and CCs. 


The present variability study provides new constraints on the star formation history of
LeoI. From a qualitative point of view, the existence of 106 RRLs, along with a well defined HB, tells us about a
very old event of star formation experienced by the galaxy more than
10 Gyr ago. Moreover, the detection of a large sample of
Cepheids, 51, suggests that an intermediate-age episode of star formation
also occurred, accordingly with a very well defined RC feature. Finally, the number of detected objects tell us that these two events of star formation
involved a considerable amount of gas.  

However, our knowledge of variable stars can drive more quantitative conclusions.
Our sample of Cepheids occupy a very intriguing and poorly
investigated luminosity and color
range, $M_V$$\lsim$--0.8 mag and 0.2 $\lsim B-V \lsim$ 0.5. Few Cepheids
that occupy this region of the CMD are observed in other dSph galaxies.
On the contrary, many of them are observed in irregular dwarfs like
LeoA \citep{dolphin02}, SextansA and B \citep{sandage85,piotto94,dolphin03},
and Phoenix \citep{gallart04}. Using current scaled--solar theoretical prescriptions, we have shown
that a very narrow range in stellar masses (around the transition mass, M$_{HeF}$) and
metallicity (--1.5$\lsim$[Fe/H]$\lsim$--1.3) account for LeoI Cepheid
properties. The mean metallicity agrees quite well with recent
spectroscopic measurements. Moreover, the quoted scenario accounts for the difference in
the Cepheid period distribution (peaked at 1.26d instead of 0.5d)
observed in LeoI when compared with Carina and Fornax dSphs.
For these galaxies, we are carrying out extensive, archival-based variability studies in order to
identify the reasons of the huge difference in their Cepheid sample.

The current scenario depends on the adopted theoretical
assumptions. In a forthcoming paper we plan to investigate the
dependency on alpha-enhancement, helium content and mass loss since
these inputs affect the luminosity and colors of central helium
burning phases \citep{castellani00,girardi01,girardi02}.

\acknowledgments
We warmly thanks S. Cassisi. GF has been supported by the INAF fellowship 2009
grant and MM from IAC (grants 310394 and 301204), the Education and
Science Ministry of Spain (grant AYA2010-16717). GB acknowledges
support from the ESO Visitor Programme.

\bibliographystyle{apj} 

\end{document}